\documentclass{iaus}
\usepackage{natbib}
\usepackage{graphicx}
\newcommand{\degree}{\ensuremath{^\circ}}

\title[Is our Sun a Singleton] 
{Is our Sun a Singleton?}

\author[Malmberg et {\sl al.}]   
{D. Malmberg$^1$ , M. B. Davies$^1$,  J.~E. Chambers$^2, $F. De Angeli 
$^3$, \break R.~P. Church$^1$, D. Mackey$^4$ \and M. I. Wilkinson$^5$}

\affiliation{$^1$Lund Observatory, Box 43, SE-221 00, Lund, Sweden \break 
email: danielm@astro.lu.se \\[\affilskip] 
$^2$Department of Terrestrial Magnetism, Carnegie Institution of Washington, 
5241 Broad Branch Road NW, Washington DC 20015, USA \\[\affilskip]
$^3$Institute of Astronomy, Madingley Road, Cambridge, CB3 OHA, 
UK \\[\affilskip]
$^4$Institute for Astronomy, University of Edinburgh, Royal Observatory, 
Blackford Hill, Edinburgh, EH9 3HJ, UK \\[\affilskip]
$^5$Department of Physics and Astronomy, University of Leicester, 
Leicester, LE1 7RH, UK \\[\affilskip]}

\pubyear{2007}
\volume{246}  
\pagerange{119--126}
\date{?? and in revised form ??}
\jname{Proceedings Title IAU Symposium}
\editors{A.C. Editor, B.D. Editor \& C.E. Editor, eds.}
\begin{document}

\maketitle

\begin{abstract}
Most stars are formed in a cluster or association, where the number density 
of stars can be high. This means that a large fraction of initially-single
stars will undergo close encounters with other stars and/or exchange
into binaries. We describe how such close encounters and exchange
encounters can affect the properties of a planetary system around
a single star. We define a singleton as a single star which has never
suffered close encounters with other stars or spent time within a
binary system. It may be that planetary systems similar to our
own solar system can only survive around singletons. Close 
encounters or the presence of a stellar companion will perturb 
the planetary system, often leaving planets on tighter and more 
eccentric orbits. Thus planetary systems which initially resembled 
our own solar system may later more closely resemble some of the 
observed exoplanet systems. 

\keywords{binaries: general, open clusters and associations: general, planetary systems: general}
\end{abstract}

Stars are most often formed in some sort of cluster or association. 
In such environments the number density of stars can be significantly
higher than in the solar neighborhood. Thus, close encounters between
stars might be common. If a single star with a planetary system suffers a close
encounter with another star, the orbits of the planets might be changed.
Sometimes this change can be enough for one or more planets to be 
ejected entirely from the system. Most likely this will happen long after
the encounter, due to the strong planet-planet interactions
induced by the close
encounter. If one or more planets are ejected, the remaining planets will most
often be left on tighter and more eccentric orbits. It is also possible that 
the close encounter does not cause  the ejection of any planets and instead
just stirs up the eccentricities of the planets somewhat.

If a single star instead encounters a binary system, it can be exchanged 
into it. When this occurs, the orientation of the orbital plane of the 
planets with respect to that of the companion star is completely random. 
This means that in about 70 per cent of the cases, the inclination between 
the two will be larger than $40\degree$. When that happens, the Kozai Mechanism
will operate \citep{1962AJ.....67..591K}. Given that the binary is not too 
wide, the Kozai Mechanism will cause the eccentricities of the planets to 
oscillate. If the planetary system contains multiple planets, this 
eccentricity pumping can cause strong planet-planet interactions, 
causing the orbits of the planets to change significantly and
sometimes also ejecting one or more planets \citep{2007MNRAS.377L...1M}

\begin{figure}
 \resizebox{8truecm}{!}{\includegraphics{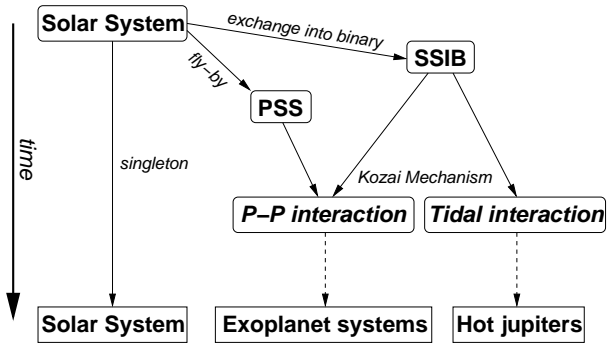}}
 \caption{This flow-diagram outlines what can happen to a solar
system orbiting an initially single star inside a stellar cluster.
SSIB stands for Solar System In Binary, PSS stands for Perturbed
Solar System and P-P interaction stands for Planet-Planet interaction.
}\label{fig:flowdiag}
\end{figure}

In our own solar system, the orbits of the planets are nearly
circular. Furthermore, all the massive planets are found far out
from the sun. This is however not the case for the
observed exoplanets. Many of these systems contain
one or more eccentric planets and also often massive planets on tight
orbits. We propose that at least some
of the observed exoplanet systems were once resembling the solar system,
but were later altered into the planetary system we observe today, 
through interactions in stellar clusters.
It might be that planetary systems like our own solar system can only 
exist around stars which formed single and has never experienced a close 
encounter with another star or been inside a stellar binary. We define
such a star to be called a singleton. In Fig. \ref{fig:flowdiag} 
we give an outline of what can happen to a solar
system like planetary system when inside a stellar cluster. Only around
singletons will a planetary system remain solar-system-like until today, while
if it orbits a star which suffer strong interactions with other stars it will
today instead either be a Hot Jupiter system 
\citep[see for example][]{2007arXiv0705.4285F,2007arXiv0706.0732W} or contain 
planets on elliptical orbits \citep{2005ApJ...627.1001T}.

To explore how large the fraction of singletons is in the solar
neighborhood (and thus to understand how common solar system like planetary
system may be) we have numerically simulated a large range of stellar clusters, 
typical of those in which most of the stars in the solar neighborhood formed 
\citep{2007MNRAS.378.1207M}.
From our simulations we estimate the singleton fraction for single stars
with masses similar to that of the sun to be between 0.90 and 0.95. This
means, that between 5 and 10 per cent of all planetary systems around
solar mass stars can have been altered by dynamical interactions in stellar
clusters, such as described above, into some of the observed exoplanet systems.

\begin{acknowledgments}
Melvyn B. Davies is a Royal Swedish Academy Research Fellow supported by a 
grant from the Knut and Alice Wallenberg Foundation. Ross P. Church is funded 
by a grant from the Swedish Institute. Dougal Mackey is supported by a Marie 
Curie Excellence Grant under contract MCEXT-CT-2005-025869. Mark Wilkinson 
acknowledges support from a Royal Society University Research Fellowship. 
\end{acknowledgments}

\end{document}